\documentclass[mathleft,eps
]{an}

\usepackage{graphicx}
\usepackage{times}
\usepackage{bm}

\begin{document}

% The following seven commands are intended for editorial usage and should be ignored by
% the author(s).
\Pagespan{1}{}% Document's page range. 
% If second parameter is left empty, the last page is computed automatically.
\Yearpublication{}%
\Yearsubmission{}%
\Month{}%   
\Volume{}%  
\Issue{}% 
\DOI{}% 

\title{Dirac and Majorana Field Operators with Self/Anti-Self Charge Conjugate States}

\author{Valeriy V. Dvoeglazov
%\inst{1}
%\fnmsep
\thanks{Corresponding author:
  \email{valeri@fisica.uaz.edu.mx}\newline}}

\titlerunning{Dirac and Majorana}
\authorrunning{V. V. Dvoeglazov}
\institute{UAF, Universidad Aut\'onoma de Zacatecas, Ap. Postal 636, Suc. 3, Zacatecas 98061 Zac., M\'exico}

\received{}
\accepted{}
\publonline{}


\abstract{We discuss relations between Dirac and Majorana-like field operators 
with self/anti-self charge conjugate states. The connections with recent models of
several authors have been found.}

%\date{\empty}

\maketitle

In Refs. (Dvoeglazov 2003,2006,2009,2011,2013,2015,\, 2016) we considered the procedure of construction of the field operators {\it ab initio} 
(including for neutral particles). \linebreak The Bogoliubov-Shirkov method has been used, Ref. (Bogoliubov \& Shirkov 1984).

In the present article we investigate the spin-1/2 case for self/anti-self  charge conjugate states.
We look for interrelations between the Dirac field operator 
and the Majorana field operator. It seems that the calculations  give mathematically and physically reasonable results in the helicity basis only.

We write the charge conjugation operator  into the form:
\begin{equation}
C= e^{i\theta_c} \pmatrix{0&0&0&-i\cr
0&0&i&0\cr
0&i&0&0\cr
-i&0&0&0\cr} {\cal K} = -e^{i\theta_c} \gamma^2 {\cal K}\,.
\end{equation}
It is the anti-linear operator of charge conjugation. ${\cal K}$ is the complex conjugation operator. We  define the {\it self/anti-self} charge-conjugate 4-spinors 
in the momentum space \linebreak (Ahluwalia 1996):
\begin{eqnarray}
C\lambda^{S,A} ({\bf p}) &=& \pm \lambda^{S,A} ({\bf p})\,,\\
C\rho^{S,A} ({\bf p}) &=& \pm \rho^{S,A} ({\bf p})\,.
\end{eqnarray}
Thus,
\begin{equation}
\lambda^{S,A} (p^\mu)=\pmatrix{\pm i\Theta \phi^\ast_L ({\bf p})\cr
\phi_L ({\bf p})}\,,
\end{equation}
and
\begin{equation}
\rho^{S,A} ({\bf p})=\pmatrix{\phi_R ({\bf p})\cr \mp i\Theta \phi^\ast_R ({\bf p})}\,.
\end{equation}
$\phi_L$, $\phi_R$ can be boosted with the Lorentz transformation \linebreak $\Lambda_{L,R}$ 
matrices.\footnote{Such definitions of 4-spinors differ, of course, from the original Majorana definition in x-representation:
\begin{equation}
\nu (x) = \frac{1}{\sqrt{2}} (\Psi_D (x) + \Psi_D^c (x))\,,
\end{equation}
$C \nu (x) = \nu (x)$ that represents the positive real $C-$ parity field operator. However, the momentum-space 
Majorana-like spinors 
open various possibilities for description of neutral  particles 
(with experimental consequences, see (Kirchbach \& Compean \& Noriega 2004).}

The rest $\lambda -$ and $\rho -$ spinors are:\footnote{The choice of the helicity parametrization for
${\bf p}\rightarrow {\bf 0}$ is doubtful in Ref. (Ahluwalia \& Grumiller 2005), and it leads to unremovable contradictions, in my opinion.}
\begin{eqnarray}
\lambda^S_\uparrow ({\bf 0}) &=& \sqrt{\frac{m}{2}}
\pmatrix{0\cr i \cr 1\cr 0}\,,\,
\lambda^S_\downarrow ({\bf 0})= \sqrt{\frac{m}{2}}
\pmatrix{-i \cr 0\cr 0\cr 1}\,,\,\\
\lambda^A_\uparrow ({\bf 0}) &=& \sqrt{\frac{m}{2}}
\pmatrix{0\cr -i\cr 1\cr 0}\,,\,
\lambda^A_\downarrow ({\bf 0}) = \sqrt{\frac{m}{2}}
\pmatrix{i\cr 0\cr 0\cr 1}\,,\,\\
\rho^S_\uparrow ({\bf 0}) &=& \sqrt{\frac{m}{2}}
\pmatrix{1\cr 0\cr 0\cr -i}\,,\,
\rho^S_\downarrow ({\bf 0}) = \sqrt{\frac{m}{2}}
\pmatrix{0\cr 1\cr i\cr 0}\,,\,\\
\rho^A_\uparrow ({\bf 0}) &=& \sqrt{\frac{m}{2}}
\pmatrix{1\cr 0\cr 0\cr i}\,,\,
\rho^A_\downarrow ({\bf 0}) = \sqrt{\frac{m}{2}}
\pmatrix{0\cr 1\cr -i\cr 0}\,.
\end{eqnarray}
Thus, in this basis the explicit forms of the 4-spinors of the second kind  $\lambda^{S,A}_{\uparrow\downarrow}
({\bf p})$ and $\rho^{S,A}_{\uparrow\downarrow} ({\bf p})$
are:
\begin{eqnarray}
\lambda^S_\uparrow ({\bf p}) &=& \frac{1}{2\sqrt{E_p+m}}
\pmatrix{ip_l\cr i (p^- +m)\cr p^- +m\cr -p_r},\\
\lambda^S_\downarrow ({\bf p}) &=& \frac{1}{2\sqrt{E_p+m}}
\pmatrix{-i (p^+ +m)\cr -ip_r\cr -p_l\cr (p^+ +m)},\nonumber\\
\lambda^A_\uparrow ({\bf p}) &=& \frac{1}{2\sqrt{E_p+m}}
\pmatrix{-ip_l\cr -i(p^- +m)\cr (p^- +m)\cr -p_r},\nonumber\\
\lambda^A_\downarrow ({\bf p}) &=& \frac{1}{2\sqrt{E_p+m}}
\pmatrix{i(p^+ +m)\cr ip_r\cr -p_l\cr (p^+ +m)},\nonumber\\
\rho^S_\uparrow ({\bf p}) &=& \frac{1}{2\sqrt{E_p+m}}
\pmatrix{p^+ +m\cr p_r\cr ip_l\cr -i(p^+ +m)},\\
\rho^S_\downarrow ({\bf p}) &=& \frac{1}{2\sqrt{E_p+m}}
\pmatrix{p_l\cr (p^- +m)\cr i(p^- +m)\cr -ip_r},\nonumber\\
\rho^A_\uparrow ({\bf p}) &=& \frac{1}{2\sqrt{E_p+m}}
\pmatrix{p^+ +m\cr p_r\cr -ip_l\cr i (p^+ +m)},\nonumber\\
\rho^A_\downarrow ({\bf p}) &=& \frac{1}{2\sqrt{E_p+m}}
\pmatrix{p_l\cr (p^- +m)\cr -i(p^- +m)\cr ip_r}.\nonumber
\end{eqnarray}
As we showed $\lambda -$ and $\rho -$ 4-spinors are not \linebreak the eigenspinors of the helicity. Moreover, 
$\lambda -$ and $\rho -$ are not (if we use the parity matrix 
\linebreak
$P=\pmatrix{0&1\cr 1&0}R$) the eigenspinors of the parity, as opposed to the Dirac case.
The indices $\uparrow\downarrow$ should be referred to the chiral helicity 
quantum number introduced 
in the 60s, $\eta=-\gamma^5 h$, for $\lambda$ spinors.
While 
\begin{equation}
Pu_\sigma ({\bf p}) = + u_\sigma ({\bf p})\,,
Pv_\sigma ({\bf p}) = - v_\sigma ({\bf p})\,,
\end{equation}
we have
\begin{equation}
P\lambda^{S,A} ({\bf p}) = \rho^{A,S} ({\bf p})\,,
P \rho^{S,A} ({\bf p}) = \lambda^{A,S} ({\bf p})
\end{equation}
for the Majorana-like momentum-space 4-spinors on \linebreak the first quantization level.
In this basis one has
\begin{eqnarray}
\rho^S_\uparrow ({\bf p}) \,&=&\, - i \lambda^A_\downarrow ({\bf p})\,,\,
\rho^S_\downarrow ({\bf p}) \,=\, + i \lambda^A_\uparrow ({\bf p})\,,\,\\
\rho^A_\uparrow ({\bf p}) \,&=&\, + i \lambda^S_\downarrow ({\bf p})\,,\,
\rho^A_\downarrow ({\bf p}) \,=\, - i \lambda^S_\uparrow ({\bf p})\,.
\end{eqnarray}
The analogs of the spinor normalizations (for $\lambda^{S,A}_{\uparrow\downarrow}
({\bf p})$ and $\rho^{S,A}_{\uparrow\downarrow} ({\bf p})$) are the following ones:
\begin{eqnarray}
\overline \lambda^S_\uparrow ({\bf p}) \lambda^S_\downarrow ({\bf p}) \,&=&\,
- i m \,,
\overline \lambda^S_\downarrow ({\bf p}) \lambda^S_\uparrow ({\bf p}) \,= \,
+ i m \,,\\
\overline \lambda^A_\uparrow ({\bf p}) \lambda^A_\downarrow ({\bf p}) \,&=&\,
+ i m \,,
\overline \lambda^A_\downarrow ({\bf p}) \lambda^A_\uparrow ({\bf p}) \,=\,
- i m \,,\\
\overline \rho^S_\uparrow ({\bf p}) \rho^S_\downarrow ({\bf p}) \, &=&  \,
+ i m\,,
\overline \rho^S_\downarrow ({\bf p}) \rho^S_\uparrow ({\bf p})  \, =  \,
- i m\,,\\
\overline \rho^A_\uparrow ({\bf p}) \rho^A_\downarrow ({\bf p})  \,&=&\,
- i m\,,
\overline \rho^A_\downarrow ({\bf p}) \rho^A_\uparrow ({\bf p}) \,=\,
+ i m\,.
\end{eqnarray}
All other conditions are equal to zero.

The $\lambda -$ and $\rho -$ spinors are connected with the $u-$ and $v-$ spinors by the  following formula:
\begin{eqnarray}
\hspace*{-10mm}\pmatrix{\lambda^S_\uparrow ({\bf p}) \cr \lambda^S_\downarrow ({\bf p}) \cr
\lambda^A_\uparrow ({\bf p}) \cr \lambda^A_\downarrow ({\bf p})\cr} = {1\over
2} \pmatrix{1 & i & -1 & i\cr -i & 1 & -i & -1\cr 1 & -i & -1 & -i\cr i&
1& i& -1\cr} \pmatrix{u_{+1/2} ({\bf p}) \cr u_{-1/2} ({\bf p}) \cr
v_{+1/2} ({\bf p}) \cr v_{-1/2} ({\bf p})\cr}\nonumber
\\ \label{connect}
\end{eqnarray}
provided that the 4-spinors have the same physical dimension.\footnote{The change of 
the mass dimension of the field operator has no sufficient foundations because the Lagrangian can be constructed on
using the coupled Dirac equations, see Ref.~(Dvoeglazov 1995). After that one can play with $\sqrt{m}$
to reproduce all possible mathematical results, which may (or may not) answer to the physical reality.}

We construct the field operators on using the Bogoliubov-Shirkov procedure with $\lambda^S_\eta (p)$:
\begin{eqnarray}
&&\Psi (x) = {1\over (2\pi)^3} \int d^4 p \,\delta (p^2 -m^2) e^{-ip\cdot x}
\Psi (p) =\nonumber\\
&=& {1\over (2\pi)^3} \sum_{\eta=\uparrow\downarrow}^{}\int d^4 p \, \delta (p_0^2 -E_p^2) e^{-ip\cdot x}
\sqrt{m} \nonumber\\
&&[\lambda^S_\eta (p_0, {\bf p}) c_\eta (p_0, {\bf p})] =\label{fo}\\
&=&{\sqrt{m}\over (2\pi)^3} \int {d^4 p \over 2E_p} [\delta (p_0 -E_p) +\delta (p_0 +E_p) ]\nonumber\\ 
&&[\theta (p_0) +\theta (-p_0) ] e^{-ip\cdot x}
\sum_{\eta=\uparrow\downarrow}^{} \lambda^S_\eta (p) c_\eta (p) \nonumber\\
&=& {\sqrt{m}\over (2\pi)^3} \sum_{\eta=\uparrow\downarrow}^{} \int {d^4 p \over 2E_p} [\delta (p_0 -E_p) +\delta (p_0 +E_p) ] 
\nonumber\\
&&\left
[\theta (p_0) (p)\lambda^S_\eta (p) c_\eta (p) e^{-ip\cdot x}  + \right.\nonumber\\
&+&\left.\theta (p_0) \lambda^S_\eta (-p) c_\eta (-p) e^{+ip\cdot x} \right ]\nonumber\\ 
&=& {\sqrt{m}\over (2\pi)^3} \sum_{\eta=\uparrow\downarrow}^{} \int {d^3 {\bf p} \over 2E_p} \theta(p_0)\nonumber\\  
&&\left [ \lambda^S_\eta (p) c_\eta (p)\vert_{p_0=E_p} e^{-i(E_p t-{\bf p}\cdot {\bf x})}  +\right.\nonumber\\
&+&\left.\lambda^S_\eta (-p) c_\eta (-p)\vert_{p_0=E_p} e^{+i (E_p t- {\bf p}\cdot {\bf x})} 
\right ]\nonumber
\end{eqnarray}

Thus, comparing with the Dirac field operator we have 1) instead of
$u_h (\pm p)$ we have $\lambda^S_\eta (\pm p)$; 2) possible change of the annihilation operators,
$a_h \rightarrow c_\eta$. Apart, one can make corresponding changes due to 
normalization factors. Thus, we should have 
\begin{equation}
\sum_{\eta=\uparrow\downarrow}^{} \lambda^A_\eta (p) d_\eta^\dagger (p) = \sum_{\eta=\uparrow\downarrow}^{} \lambda^S_\eta (-p) c_\eta (-p)\,.\label{dcop1}
\end{equation}
Multiplying by $\overline\lambda^A_{-\kappa} (p)$ or $\overline\lambda^S_{-\kappa} (-p)$, respectively, 
we find surprisingly:
\begin{eqnarray}
d^\dagger_\kappa (p)&=& -\frac{ip_y}{p}\sigma^y_{\kappa\tau}  c_\tau (-p)\,,\\
c_\kappa (-p)&=& -\frac{ip_y}{p}\sigma^y_{\kappa\tau}  d^\dagger_\tau (p)\,.
\end{eqnarray}
The above-mentioned contradiction may be related to the possibility of the conjugation which is different from that of Dirac.
Both in the Dirac-like case and the Majorana-like case ($c_\eta (p)= e^{-i\varphi} d_\eta (p)$) we have difficulties in the construction of field operators (Dvoeglazov 2018b).

The bi-orthogonal anticommutation relations are given in Ref.~(Ahluwalia 1996). See other details in \linebreak
Ref.~(Dvoeglazov 1995a, 1997).
Concerning with the $P$,$C$ and $T$ properties of the corresponding states \linebreak see Ref.~(Dvoeglazov 2011) in this model.

Similar formulations have been  presented in \linebreak Refs.~(Markov 1937), 
and~(Barut \& Ziino 1993). Namely, the reflection properties are different for some solutions 
of relativistic equations therein. Two opposite signs at the mass terms have been taken into account.
The group-theoretical basis for such doubling has been  given
in the papers by Gelfand, Tsetlin (1957) and Sokolik~(1957), who first presented 
the theory of 5-dimensional spinors (or, the one in the 2-dimensional projective representation of the inversion group) in 1956 
(later called as ``the Bargmann-Wightman-Wigner-type quantum field theory" in 1993). \linebreak The corresponding connection 
with the time reversion has been clarified therein. It was one of the first attempts to explain the $K$-meson decays.
M. Markov proposed  {\it two} Dirac equations with opposite signs at the mass term~(Markov 1937) to be taken into account:
\begin{eqnarray}
\left [ i\gamma^\mu \partial_\mu - m \right ]\Psi_1 (x) &=& 0\,,\\
\left [ i\gamma^\mu \partial_\mu + m \right ]\Psi_2 (x) &=& 0\,.
\end{eqnarray}
In fact, he studied all properties of this relativistic quantum model (while the quantum
field theory has not yet been completed in 1937). Next, he added and  subtracted these equations. What did he obtain?
\begin{eqnarray}
i\gamma^\mu \partial_\mu \varphi (x) - m \chi (x) &=& 0\,,\\
i\gamma^\mu \partial_\mu \chi (x) - m \varphi (x) &=& 0\,.
\end{eqnarray}
Thus, the corresponding $\varphi$ and $\chi$ solutions can be presented as some superpositions of the Dirac 4-spinors $u-$ and $v-$.
These equations, of course, can be identified with the equations for the Majorana-like $\lambda -$ and $\rho -$, which we presented 
in Ref.~(Dvoeglazov 1995b).\footnote{Of course, the signs at the mass terms
depend on, how do we associate the positive- or negative- frequency solutions with $\lambda$ and $\rho$.}
\begin{eqnarray}
i \gamma^\mu \partial_\mu \lambda^S (x) - m \rho^A (x) &=& 0 \,,
\label{11}\\
i \gamma^\mu \partial_\mu \rho^A (x) - m \lambda^S (x) &=& 0 \,,
\label{12}\\
i \gamma^\mu \partial_\mu \lambda^A (x) + m \rho^S (x) &=& 0\,,
\label{13}\\
i \gamma^\mu \partial_\mu \rho^S (x) + m \lambda^A (x) &=& 0\,.
\label{14}
\end{eqnarray}
Neither of them can be regarded as the Dirac equation.
However, they can be written in the 8-component form as follows:
\begin{eqnarray}
\left [i \Gamma^\mu \partial_\mu - m\right ] \Psi_{_{(+)}} (x) &=& 0\,,
\label{psi1}\\
\left [i \Gamma^\mu \partial_\mu + m\right ] \Psi_{_{(-)}} (x) &=& 0\,,
\label{psi2}
\end{eqnarray}
with
\begin{eqnarray}
&&\Psi_{(+)} (x) = \pmatrix{\rho^A (x)\cr
\lambda^S (x)\cr},
\Psi_{(-)} (x) = \pmatrix{\rho^S (x)\cr
\lambda^A (x)\cr}\,,\\
&&\Gamma^\mu =\pmatrix{0 & \gamma^\mu\cr
\gamma^\mu & 0\cr}.
\end{eqnarray}
It is possible to find the corresponding Lagrangian, projection operators, and the Feynman-Dyson-Stueckelberg propagator.
For example,
\begin{eqnarray}
{\cal L} &=&\frac{i}{2} \left . [\overline\Psi_{(+)} \Gamma^{\mu}\partial_\mu \Psi_{(+)} - (\partial_\mu \overline\Psi_{(+)}) \Gamma^{\mu}
\Psi_{(+)} + \right . \nonumber\\
&+&\left . \overline\Psi_{(-)} \Gamma^{\mu}\partial_\mu \Psi_{(-)} - (\partial_\mu \overline\Psi_{(-)}) \Gamma^{\mu}
\Psi_{(-)} \right ] -\nonumber\\
&-& m [\overline\Psi_{(+)}\Psi_{(+)} - \overline\Psi_{(-)}\Psi_{(-)}]\,.
\end{eqnarray}
The projection operator $P_+$ can be easily found, as usual,
\begin{equation}
P_+ = \frac{\Gamma_\mu p^\mu +m}{2m}\,.
\end{equation}
However, due to the fact that $P_-$ satisfies the Dirac equation with the opposite sign, we cannot have $P_+ + P_- =1$.
This is not surprising  because the corresponding states $\Psi_\pm$ do not form the complete system of the 8-dimensional space.
One should consider the states $\Gamma_5 \Psi_\pm ({\bf p})$ too.
See also~(Dvoeglazov 2018a) for the methods of obtaining the propagators in the non-trivial cases.

In the previous papers I explained: the connection with the Dirac spinors has 
been found~(Dvoeglazov 1995b; \linebreak Kirchbach \& Compean \& Noriega 2004) through the unitary matrix, 
provided that the 4-spinors have the same physical dimension.\footnote{The reasons of the change of the fermion 
mass dimension are unclear in the recent works on {\it elko}.}
Thus, this represents
itself the rotation of the spin-parity basis. However, it is usually assumed that the $\lambda-$ and $\rho-$ spinors describe the neutral particles,
meanwhile, the $u-$ and $v-$ spinors describe the charged particles. Kirchbach, Compean and Noriega (2004) found the amplitudes for 
neutrinoless double beta decay ($00\nu\beta$) in this scheme. It is obvious from (\ref{connect}) that there are some additional terms comparing with the standard calculations of those amplitudes. 
One can also re-write the above equations into the two-component forms. Thus, one obtains the Feynman and Gell-Mann (1958) equations.

Barut and Ziino~(1993) proposed yet another model. \linebreak They considered
$\gamma^5$ operator as the operator of the charge conjugation. In their case the self/anti-self charge conjugate states
are, at the same time, the eigenstates of the chirality. Thus, the charge-conjugated
Dirac equation has a different sign compared with the ordinary formulation:
\begin{equation}
[i\gamma^\mu \partial_\mu + m] \Psi_{BZ}^c =0\,,
\end{equation}
and the so-defined charge conjugation applies to the whole system,  fermion + electro\-magnetic field, $e\rightarrow -e$
in the covariant derivative. The superpositions of the $\Psi_{BZ}$ and $\Psi_{BZ}^c$ also give us 
the ``doubled Dirac equation", as the equations for $\lambda-$ and $\rho-$ spinors. 
The concept of the doubling of the Fock space has been
developed in the Ziino works, cf. (Gelfand \& Tsetlin 1957; Sokolik 1957; Dvoeglazov 1998) in the framework of the quantum field theory~(Ziino 1996). 
Next, it is interesting to note that we have for the Majorana-like field operators ($a_\eta ({\bf p}) = b_\eta ({\bf p})$):
\begin{eqnarray}
\lefteqn{\left [ \nu^{^{ML}} (x^\mu) + {\cal C} \nu^{^{ML\,\dagger}} (x^\mu) \right ]/2 = 
\int {d^3 {\bf p} \over (2\pi)^3 } {1\over 2E_p} }\\
&&\sum_\eta \left [ \pmatrix{i\Theta \phi_{_L}^{\ast \, \eta} (p^\mu) \cr 0\cr} a_\eta
(p^\mu)  e^{-ip\cdot x} +\right.\nonumber\\
&+&\left. \pmatrix{0\cr
\phi_L^\eta (p^\mu)\cr } a_\eta^\dagger (p^\mu) e^{ip\cdot x} \right ]\,,\nonumber
\\
\lefteqn{\left [\nu^{^{ML}} (x^\mu) - {\cal C} \nu^{^{ML\,\dagger}} (x^\mu) \right
]/2 = \int {d^3 {\bf p} \over (2\pi)^3 } {1\over 2E_p}} \\
&&\sum_\eta \left
[\pmatrix{0\cr \phi_{_L}^\eta (p^\mu) \cr } a_\eta (p^\mu)  e^{-ip\cdot x} +\right.\nonumber\\
&+&\left. \pmatrix{-i\Theta \phi_{_L}^{\ast\, \eta} (p^\mu)\cr 0
\cr } a_\eta^\dagger (p^\mu) e^{ip\cdot x} \right ]\, . \nonumber
\end{eqnarray}
This naturally leads to the Ziino-Barut scheme of massive chiral
fields. See, however, the recent paper~(Dvoeglazov 2018b) which deals with the problems 
of the Majorana field operator.

%\medskip

\acknowledgements
I acknowledge discussions with colleagues at recent conferences.
I am grateful to the Zacatecas University for professorship.

\newpage%%%%%%%%%%%%%%%%%%%%%%%%%%%%%%%

%\bibliography{IWARAAstronomische}%

\end{document}